\documentclass[%
 reprint,
 amsmath,amssymb,
 aps,
]{revtex4-2}

\usepackage{graphicx}
\usepackage{dcolumn}
\usepackage{bm, braket, float, bbold, mathtools}
\usepackage{hyperref}
\hypersetup{
    colorlinks=true,
    linkcolor=blue,
    urlcolor=blue,
    citecolor=blue
}

\begin{document}


\title{Grover's search algorithm for $n$ qubits with optimal number of iterations}

\author{Simanraj Sadana}
 \affiliation{Light and Matter Physics Department, Raman Research Institute, Bangalore}
 \email{simanraj@rri.res.in\\simanrajsadana@gmail.com}

\date{\today}

\begin{abstract}
The success probability of a search of $M$ targets from a database of size $N$, using Grover’s search algorithm depends critically on the number of iterations of the composite operation of the oracle followed by Grover’s diffusion operation. Although the required number of iterations scales as $\mathcal{O}(\sqrt{N})$ for large $N$, the asymptote is not a good indicator of the optimal number of iterations when $\sqrt{M/N}$ is not small. A scheme for the determination of the exact number of iterations, subject to a threshold set for the success probability of the search (probability of detecting the target state(s)), is crucial for the efficacy of the algorithm. In this work, a general scheme for the construction of $n$-qubit Grover’s search algorithm with $1 \leq M \leq N$ target states is presented, along with the procedure to find the optimal number of iterations for a successful search. It is also shown that for given $N$ and $M$, there is an upper-bound on the success probability of the algorithm.
\end{abstract}

\maketitle

\section{Introduction}
Grover’s search algorithm \cite{Grover1996} solves the problem of searching a database of size $N$, for a given target, with quadratic speed-up over any known classical algorithm for the same purpose. A typical search is performed by evaluating a function (often called the oracle), that gives a particular value for the target and another value for all the other objects in the database. A classical algorithm needs $\mathcal{O}(N)$ evaluations of the oracle to complete the search, whereas Grover’s algorithm needs $\mathcal{O}(\sqrt{N})$ evaluations (thereby giving a quadratic speed-up). 

The practical implementation of Grover’s algorithm depends critically on the number of evaluations, say $k$, of the oracle function. The prescribed formula for the optimal number of evaluations is 
\begin{align}
\label{eq:prescribedkM}
    k = \mathrm{round}\left(\frac{\pi}{4}\sqrt{\frac{N}{M}} - \frac{1}{2}\right)
\end{align} 
for a search of $M$ targets in a database of size $N$ \cite{nielsen_book, kaye_book}. However, the prescribed number of iterations, as in Eq.~\eqref{eq:prescribedkM} does not guarantee sufficient amplification of the target(s) needed for a successful search. Especially, in the case of multiple targets, Eq.~\eqref{eq:prescribedkM} may not be the optimal number iterations for a successful search if $\sqrt{M/N}$ is not small enough. 

In this work, a scheme for an $n$-qubit ($n \geq 1$) quantum circuit implementation of Grover’s search algorithm with multi-target search (without ancillae), is presented. The implementations of the algorithm for upto 3-qubit systems are demonstrated in \cite{Figgatt2017, Mandviwalla2019, abhijith2018quantum}. The key to ensuring a successful search in the case of higher-dimensions with multiple targets, is to determine the optimal number of iterations $k$, subject to a threshold set on the success probability (probability of detecting the target state(s)). The threshold on the success probability cannot be set arbitrarily close to unity in every case, as for some search problems, depending on the values of $N$ and $M$, the maximum success probability that can be achieved has an upper-bound which is less than unity. Moreover, in some cases, this upper-bound is such that Grover’s algorithm does not amplify the target states at all, and examples of such cases are presented
here. For brevity, Grover's algorithm to search a database of size $N$, for $1 \leq M \leq N$ targets is denoted by GSA$(N,M)$.

The outline of the paper is as follows. \S\ref{sec:background} gives a brief outline of the search problem and Grover's search algorithm. In \S\ref{sec:circuit_construction}, the general scheme for constructing the oracle and Grover's diffusion operator for GSA$(N,M)$ is given. The conditional for choosing the optimal number of iterations and its importance for executing a successful search is discussed in \S\ref{sec:optimalk}. The same section also highlights the existence of the upper-bound of maximum success probability of the search algorithm, along with the discussion of cases in which the algorithm does not amplify the target states. Examples of single target and multi-target Grover's search are presented in \S\ref{sec:examples}, using the quantum computing platform made available by IBM's Quantum Experience \cite{ibm_experience}. The examples also present the cases in which the optimal number of iterations is not as in Eq.~\eqref{eq:prescribedkM}, and is therefore determined numerically using the procedure discussed in \S\ref{sec:optimalk}.

\section{Background: a brief outline of Grover's algorithm}\label{sec:background}
\begin{figure}
    \includegraphics{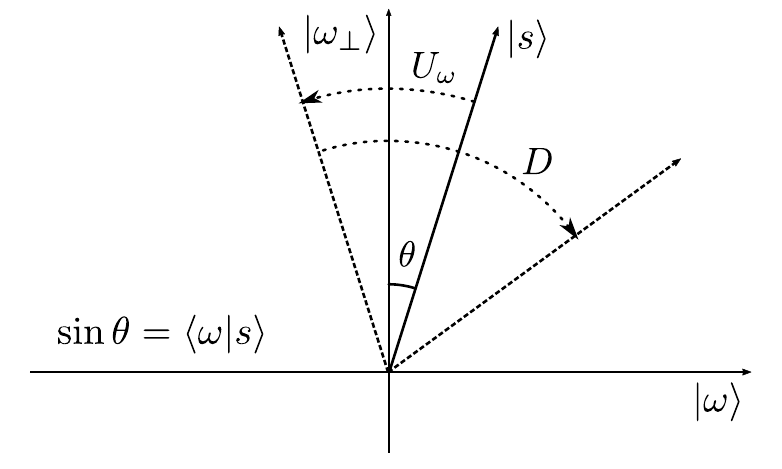}
    \caption{Geometric representation of state evolution after one iteration of oracle followed by Grover's diffusion operation.}
    \label{fig:vector_grover}
\end{figure}

In Grover’s algorithm, the objects of the database are encoded as orthogonal quantum states. For example, if $N=2^n$, then each object can be uniquely encoded as one of the computational basis states in an $n$-qubits Hilbert space. Then, assuming maximum uncertainty about the target state, the initial state is chosen to be the equal superposition of all the basis states, i.e.,
\begin{align}
    \ket{s} = H^{\otimes n}\ket{0}^{\otimes n}.
\end{align}
where $H$ is the Hadamard transformation. Grover’s algorithm amplifies the target state by applying a series of unitary operations, such that, upon a projective measurement on the final state, the target state is detected with high probability. The said unitary operation is a sequential operation of two unitary operators, i.e., the oracle followed by the diffusion operator. The unitary implementation of the oracle is
\begin{align}
    \label{eq:oracle}
    U_\omega =& \mathbb{1} - 2\ket{\omega}\bra{\omega}
\end{align}
where $\ket{\omega}$ denotes the target state (or an equal superposition of multiple target states). The oracle operation flips the sign of the target state, leaving all the non-target states unchanged. Grover’s diffusion operator is defined as
\begin{align}
    \label{eq:diffusion}
    D =& 2\ket{s}\bra{s} - \mathbb{1}
\end{align}
which reflects the state about $\ket{s}$. A repetition of this composite operation gradually amplifies the target state. A geometric representation of the state-evolution after the first iteration is shown in figure \ref{fig:vector_grover}.
In the figure, $\ket{\omega_\perp}$ is the equal superposition of all the non-target states, which is naturally orthogonal to $\ket{\omega}$. The initial state for the algorithm is $\ket{s}$ which makes an angle $\theta$ with $\ket{\omega_\perp}$ and therefore, 
\begin{align}
    \label{eq:theta_def}
    \sin\theta = \braket{\omega|s}
\end{align}
is the probability of getting the target state $\ket{\omega}$, initially. After each iteration, the state is rotated by an angle $2\theta$ and therefore, after $k$ iterations, the final state makes an angle 
\begin{align}
    \label{eq:thetak}
    \theta_k = \theta(1+2k)
\end{align}
with $\ket{\omega_\perp}$. Grover's algorithm solves the search problem by repeating the operation $D U_\omega$ until the final state is sufficiently close to $\ket{\omega}$.

\section{Construction of the circuit}\label{sec:circuit_construction}
In this section we discuss the construction of a quantum circuit for a general GSA$(N,M)$ with $n$ qubits (such that $N=2^n$). The first step is to prepare the initial state $\ket{s}$. In a typical quantum computer, the qubits are initialized to $\ket{0}$. Therefore, the state $\ket{s}$ can be prepared by applying a Hadamard gate to each of the $n$ qubits employed for the computation. 

\begin{figure}
    \includegraphics[width=1\hsize]{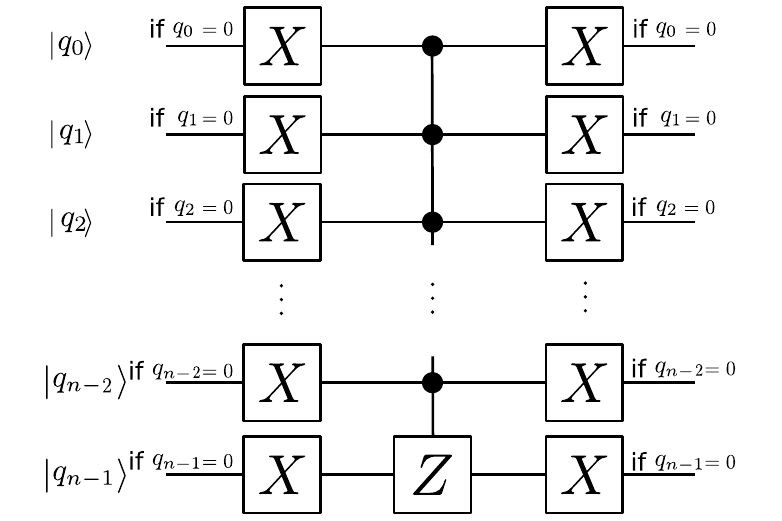}
    \caption{The general scheme for quantum circuit implementation of the oracle for a single target. The oracle for multiple targets can be constructed by concatenating the oracles for all targets.}
    \label{fig:oracle}
\end{figure}

We now construct the oracle operation for a single $n$-qubit target state. Let $\ket{q_{n-1}, \cdots, q_0}$ be the target state, where $q_i \in \{0,1\},~\forall~i$. According to Eq.~\eqref{eq:oracle}, the oracle must flip the sign of this state while leaving all the other orthogonal states unchanged. A general scheme to construct such a circuit is shown in figure \ref{fig:oracle} where an $X$-gate is applied to qubit $\ket{q_i}$ if $q_i = 0$ in the target state $\ket{q_{n-1}, \cdots, q_0}$. After the $X$-gates are applied to the appropriate qubits, the target state is (uniquely) transformed to $\ket{1}^{\otimes n}$. Then a $(n-1)$-control $Z$-gate flips the sign of $\ket{1}^{\otimes n}$. Finally, the $X$-gates are applied again to the appropriate qubits to reverse the transformation by the first set of $X$-gates. The net effect of this circuit is to flip the sign of only the target state. For multiple targets, the circuit for oracle can be constructed by concatenating the oracle corresponding to each target state. 

\begin{figure}
    \includegraphics[width=1\hsize]{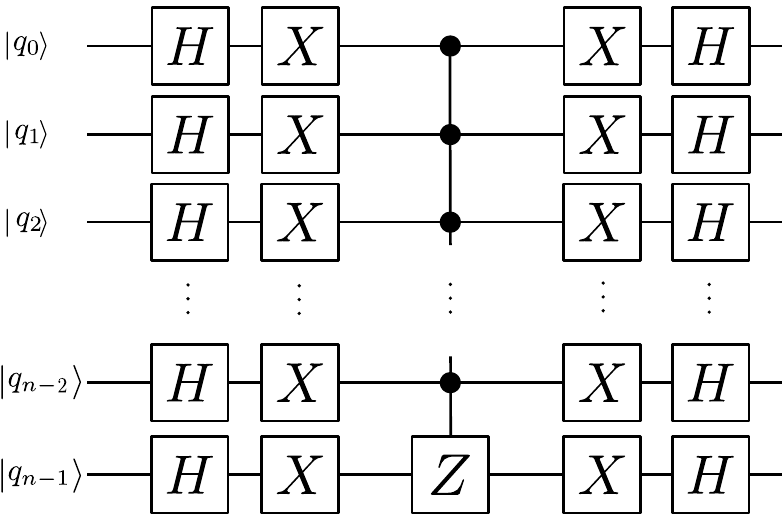}
    \caption{Quantum circuit for $n$-qubit Grover's diffusion operation with a global $\pi$-phase.}
    \label{fig:mdiffusion}
\end{figure}

After the oracle comes Grover's diffusion operator as defined in Eq.~\eqref{eq:diffusion}. The construction of an $n$-qubit diffusion circuit is simple if we redefine the diffusion operator as
\begin{align}
    \label{eq:mdiffusion}
    D \coloneqq \mathbb{1} - 2 \ket{s}\bra{s}
\end{align}
which is the diffusion operation in Eq.~\eqref{eq:diffusion} with a global phase of $\pi$. The use of Eq.~\eqref{eq:mdiffusion} instead of Eq.~\eqref{eq:diffusion} as the diffusion operator only multiplies a global phase of $\pi$ to the final state with every iteration, which does not affect the outcome of the search algorithm. Equation \eqref{eq:mdiffusion} can be expanded as
\begin{align}
    D =& H^{\otimes n} \left(\mathbb{1} - 2 \ket{0}\bra{0}\right) H^{\otimes n}
\end{align}
and the term $\mathbb{1} - 2 \ket{0}\bra{0}$ can be implemented by using the scheme for the oracle with target state set to $\ket{0}^{\otimes n}$. Figure \ref{fig:mdiffusion} shows the circuit for a diffusion operator with any number of qubits.

To complete the circuit for a given GSA$(N,M)$, the composite operation $D U_\omega$ must be repeated (by serialization) some $k$ number of times, to amplify the target states. The success probability of GSA is not a monotonic function of the number of iterations. Therefore, choosing an arbitrarily high number of iterations is not a good strategy. The determination of the optimal number of iterations is thus a crucial step in designing the circuit for GSA. The $k$ given by Eq.~\eqref{eq:prescribedkM} for any GSA$(N,M)$, is not always optimal, as discussed below.

\section{Optimal number of iterations}\label{sec:optimalk}
Let us define the \emph{optimal} number of iterations in Grover's search algorithm, as the minimum number of iterations that yields a success probability of search more than or equal to a preset threshold, say $\delta \in (0,1]$. Mathematically,
\begin{align}
    \sin^2 \left(\theta(1 + 2k_\mathrm{opt})\right) \geq \delta
\end{align}
where $k_\mathrm{opt}$ denotes the optimal number of iterations. To find $k_\mathrm{opt}$, we first maximize $\sin^2\left(\theta(1 + 2k) \right),~ \forall~ k \in \mathbb{R}$, where $\mathbb{R}$ is the set of real numbers. The maxima exist at
\begin{align}
    \label{eq:optimalk}
    k_p = (2p+1)\frac{\pi}{4 \theta} - \frac{1}{2}
\end{align}
where $p$ is a positive integer. In general, $k_p$ will not be an integer, and therefore, we must choose $\lfloor k_p \rfloor$ or $\lceil k_p \rceil$ as the number of iterations for a chosen value of $p$.

It is a natural tendency to choose $p=0$ to get the minimum number of iterations. However, such a choice does not always yield a success probability above the chosen threshold $\delta$. To substantiate this claim qualitatively, we refer to the geometric representation of state-evolution shown in figure \ref{fig:vector_grover}. With each iteration, the state-vector rotates by angle $2\theta$. If $\theta$ is very small, then after $k_0$ iterations the final angle $\theta_k$ gets very close to $\pi/2$. On the other hand, if $\theta$ is not small, then the final angle may not get sufficiently close to $\pi/2$ after just $k_0$ iterations. In that case, a different value of  $p$ must be chosen. Especially in the case of multi-target search GSA$(N,M)$, the value of $\theta$ may not be small enough depending on the ratio $M/N$. The required value of $p$ can be found numerically.

\subsection{For $M \neq N/2$ targets}
For GSA$(N,M)$ with $M\neq N/2$, 
\begin{align}
    \sin\theta = \sqrt{\frac{M}{N}}.
\end{align}
The required value of $p$ is the smallest positive integer that yields a success probability more than or equal to $\delta$, i.e.,
\begin{align}
    \label{eq:success_rate}
    p_\mathrm{opt} = \min \left\{p:\max \left(\sin^2\theta_{\lfloor k_p \rfloor}, \sin^2\theta_{\lceil k_p \rceil}\right) \geq \delta\right\}
\end{align}
Finally, the optimal number of iterations is chosen using the conditional
\begin{align}
\label{eq:kopt}
    k_\mathrm{opt} = \begin{cases}
    \lfloor k_{p_\mathrm{opt}} \rfloor,~ \mathrm{if}~ \sin^2\theta_{\lfloor k_{p_\mathrm{opt}} \rfloor} \geq \delta \\
    \lceil k_{p_\mathrm{opt}} \rceil,~ \mathrm{otherwise}
    \end{cases}.
\end{align}
The conditional in Eq.~\eqref{eq:kopt} ensures that we choose the minimum number of iterations for a successful search.
This process of determining optimal number of iterations can be executed before constructing the circuit, and the value of $p_\mathrm{opt}$ is unique for given $N$ and $M$, irrespective of the choice of target states. Therefore, determining the optimal number of iterations does not add to the complexity of the search algorithm.

\subsection{The special case of $M = N/2$}
Grover's search algorithm is ineffective if the number of search targets is $M=N/2$. In the special case of $M=N/2$,
\begin{align}
    \theta = \sin^{-1} \left(\frac{1}{\sqrt{2}}\right) = \frac{\pi}{4}.
\end{align}
From Eq.~\eqref{eq:thetak} we see that after $k$ iterations, the final state makes an angle
\begin{align}
    \theta_k = \frac{\pi}{4}(1 + 2k)
\end{align}
with $\ket{\omega_\perp}$, which is an odd multiple of $\pi/4$. Therefore, for any $k$, the success probability is 
\begin{align}
    \sin^2\left(\frac{\pi}{4}(1+2k)\right) = \frac{1}{2}
\end{align}
which means that getting a target state is equally likely to getting the non-target states. Therefore, irrespective of the number of iterations, Grover's algorithm does not amplify the target states.

\subsection{Upper-bound on success probability}
In general, we cannot set the threshold on success probability arbitrarily close to one. This happens when $\theta$ is a rational multiple of $\pi$. To prove this, suppose that 
\begin{align}
    \theta = \frac{a}{b}\pi, ~~~~ a,b~\in~ \mathbb{Z}
\end{align}
such that gcd$(a,b)=1$ then, after $k$ iterations, the final state makes an angle 
\begin{align}
    \theta_k = \theta + \frac{2ka}{b}\pi
\end{align}
with $\ket{\omega_\perp}$. Consequently, the success probability after $k$ iterations is
\begin{align}
    \sin^2\theta_k = \sin^2\left(\theta + \frac{2ka}{b} \pi\right)
\end{align}
which has a finite set of distinct values that belong to the set 
\begin{align}
    S = \left\{\sin^2\left(\theta + \frac{2ka}{b} \pi\right): 0 \leq k \leq b-1,~ k\in\mathbb{Z}\right\}
\end{align}
Therefore, the maximum success probability that can be achieved is $\max\limits_k\{S\}$. Note that GSA$(N, N/2)$ is a special case of this problem, where the set $S$ has only one element, i.e., $1/2$.

For example, for a two-qubit system with three targets, i.e., GSA$(4, 3)$, $\theta=\pi/3$ which is a rational multiple of $\pi$. The discrete set of possible success probabilities for this case is 
\begin{align}
    S_{4,3} = \left\{\sin^2\left(\frac{\pi}{3}\right), \sin^2\left(\pi\right)\right\} = \left\{\frac{3}{4}, 0\right\}
\end{align}
and therefore, the maximum success probability that can be achieved in this case is $75\%$. Note that in this specific example, Grover's search algorithm does not amplify the target states. 
\section{Examples}\label{sec:examples}
In the examples presented here, the circuit for Grover's search algorithm has been constructed and executed using the open-source quantum computing platform made available by IBM Quantum Experience (also known as IBM Qiskit). Specifically, the circuit has been run on the ``qasm\_simulator'' which assumes ideal qubits, and no decoherence. The code is available on public repository mentioned in \S\ref{sec:additional}.

\subsection{Single target search}
\subsubsection{4-qubit 1 target}
Consider a database of size $N=2^4=16$ with $M=1$ targets to search for, i.e., GSA$(16,1)$. Before constructing the circuit, we calculate the optimal number of iterations. Using Eq.~\eqref{eq:theta_def} we get
\begin{align}
    \theta = \sin^{-1}\left(\frac{1}{4}\right)
\end{align}
To get the optimal number of iterations, we must set a threshold on the success probability. Then using Eqs.~\eqref{eq:optimalk}, \eqref{eq:success_rate} and \eqref{eq:kopt} we find the values of $p$ and $k_p$ numerically. Table \ref{tab:optimal_k_single} shows the values.
{\renewcommand{\arraystretch}{1.5}
\begin{table}[H]
    \centering
    \begin{tabular}{>{\centering\arraybackslash}m{0.2\hsize} >{\centering\arraybackslash}m{0.2\hsize} >{\centering\arraybackslash}m{0.2\hsize}}
         $\delta$ & $p$ & $k_p$ \\
         \hline
         0.95 & 0 & 3 \\
         0.99 & 1 & 9 \\
         0.999 & 2 & 15
    \end{tabular}
    \caption{The value of $p$ and $k_p$ for a given success threshold $\delta$ for GSA$(16,1)$.}
    \label{tab:optimal_k_single}
\end{table}}

\begin{figure}
    \includegraphics[width=1\hsize]{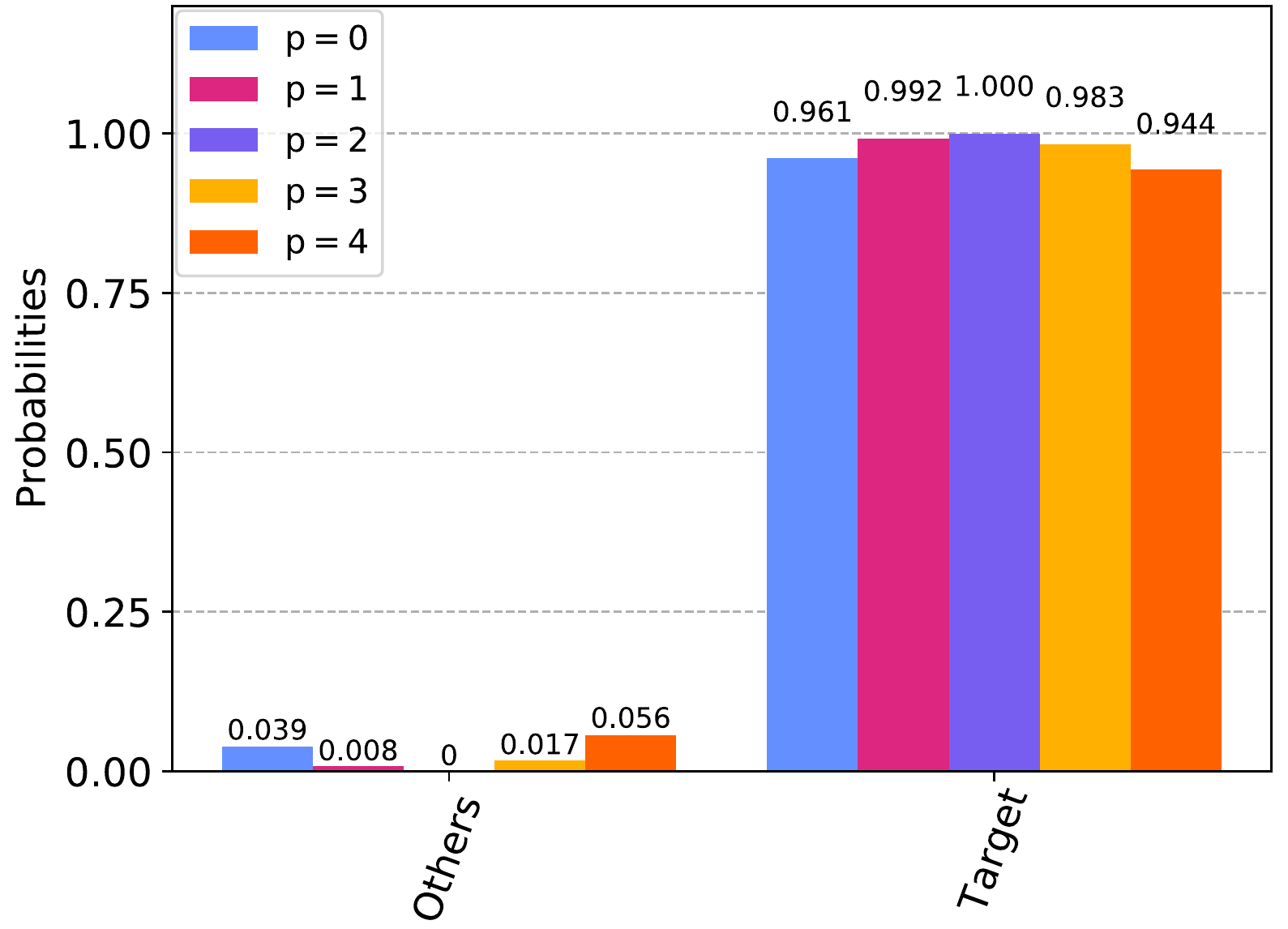}
    \caption{Result of Grover's search algorithm for GSA$(16,1)$ for different values of $p$. Here ``Target'' refers to the target state, and ``Others'' refer to all the non-target states (the probabilities of all the non-target states combined).}
    \label{fig:outcome_1101}
\end{figure}

Now, for this example, let the target state be $\ket{1101}$ (the choice of target state is irrelevant). The oracle function for this case is constructed using the scheme shown in figure \ref{fig:outcome_1101}. The circuit for a 4-qubit Grover's diffusion operator is a 4-qubit version of the circuit in figure \ref{fig:mdiffusion}. The results of the algorithm are shown in figure \ref{fig:outcome_1101}. For $p=0,1,2$ the corresponding success probabilities are above the thresholds  in table \ref{tab:optimal_k_single}. Note the the success probability does not necessarily increase with increasing value of $p$ as is evident from the outputs of the circuits with $p=3,4$. Therefore, choosing an arbitrary high number of iterations is not a good strategy. This makes the scheme for choosing optimal number of iterations, presented here, crucial for sufficient amplification of the target state(s). 

\subsubsection{7-qubits 1 target}
\begin{figure}
    \includegraphics[width=1\hsize]{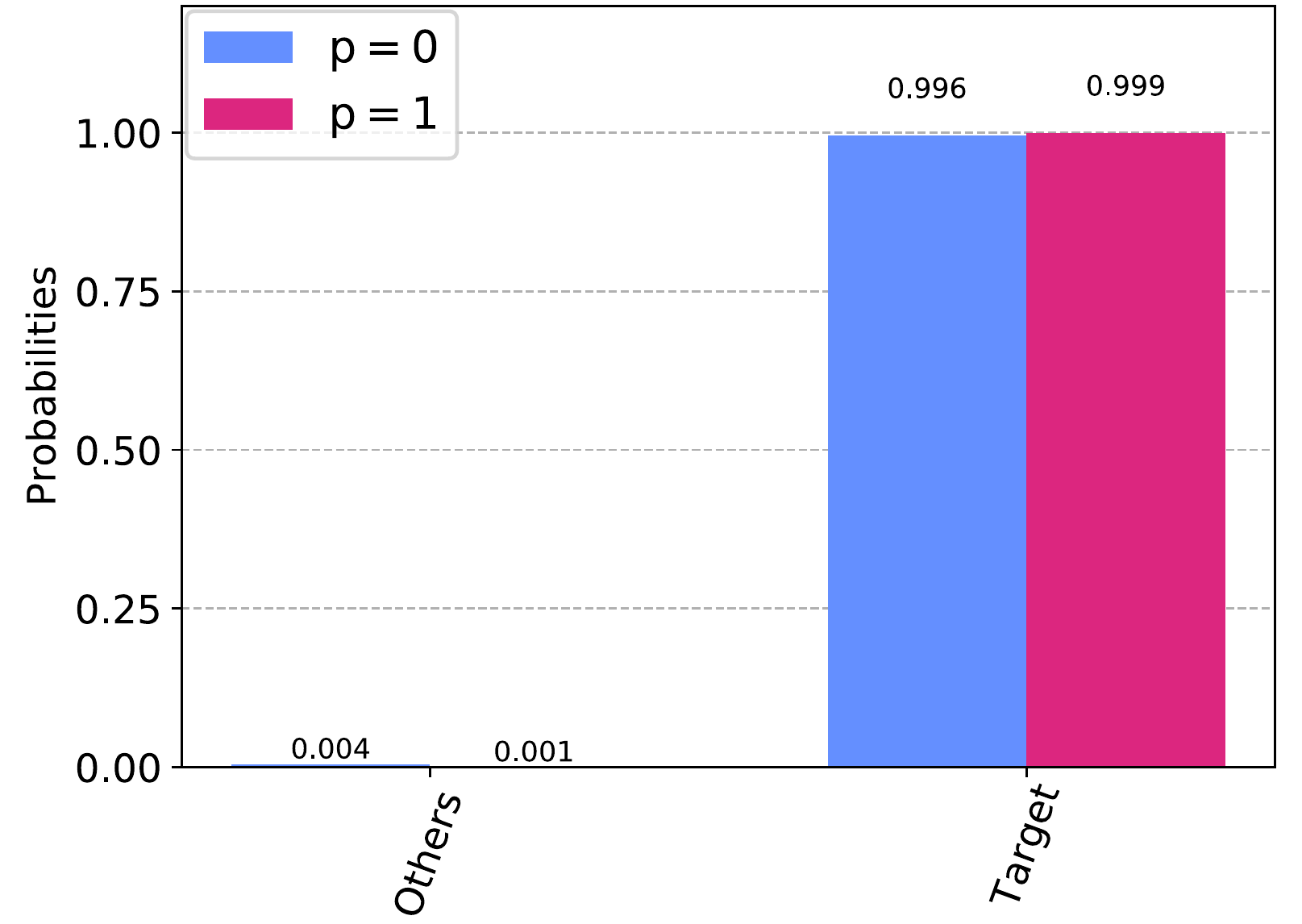}
    \caption{Result of Grover's search algorithm for GSA$(128,1)$ for different values of $p$. Here ``Target'' refers to the target state, and ``Others'' refer to all the non-target states (the probabilities of all the non-target states combined).}
    \label{fig:outcome_7_1_targets}
\end{figure}

To demonstrate the efficacy of the scheme presented here, for implementation of Grover's algorithm with higher number of qubits, a $7$-qubit version with single target search is shown here, i.e., GSA$(128, 1)$. The value of $\theta$ for this search is
\begin{align}
    \theta = \sin^{-1}\sqrt{\frac{1}{128}}.
\end{align}
The thresholds of success probability and the corresponding values of $p$ and $k_p$ are shown in table \ref{tab:optimal_k_single_7}, and the output of the circuit is shown in figure \ref{fig:outcome_7_1_targets}.
{\renewcommand{\arraystretch}{1.5}
\begin{table}[H]
    \centering
    \begin{tabular}{>{\centering\arraybackslash}m{0.2\hsize} >{\centering\arraybackslash}m{0.2\hsize} >{\centering\arraybackslash}m{0.2\hsize}}
         $\delta$ & $p$ & $k_p$ \\
         \hline
         0.95 & 0 & 8 \\
         0.99 & 0 & 8 \\
         0.999 & 1 & 26
    \end{tabular}
    \caption{The value of $p$ and $k_p$ for a given success threshold $\delta$ for GSA$(128,1)$.}
    \label{tab:optimal_k_single_7}
\end{table}}

\subsection{Multiple targets}
\subsubsection{4-qubits 9 targets}
In this example, we consider a database of size $N=2^4=16$ and $M=9$ targets, i.e., GSA$(16,9)$. The choice of targets is irrelevant. Using Eq.~\eqref{eq:theta_def},
\begin{align}
    \theta = \sin^{-1}\left(\frac{3}{4}\right).
\end{align}
Now, using Eqs.~\eqref{eq:optimalk} and \eqref{eq:success_rate} we calculate the values of $p$ and $k_p$ for different thresholds for success probability, which is shown in table \ref{tab:optimal_k_9}.
{\renewcommand{\arraystretch}{1.5}
\begin{table}[H]
    \centering
    \begin{tabular}{>{\centering\arraybackslash}m{0.2\hsize} >{\centering\arraybackslash}m{0.2\hsize} >{\centering\arraybackslash}m{0.2\hsize}}
         $\delta$ & $p$ & $k_p$ \\
         \hline
         0.95 & 2 & 4 \\
         0.99 & 3 & 6 \\
         0.999 & 3 & 6
    \end{tabular}
    \caption{The value of $p$ and $k_p$ for a given success threshold $\delta$ for GSA$(16,9)$.}
    \label{tab:optimal_k_9}
\end{table}}

\begin{figure}
    \includegraphics[width=1\hsize]{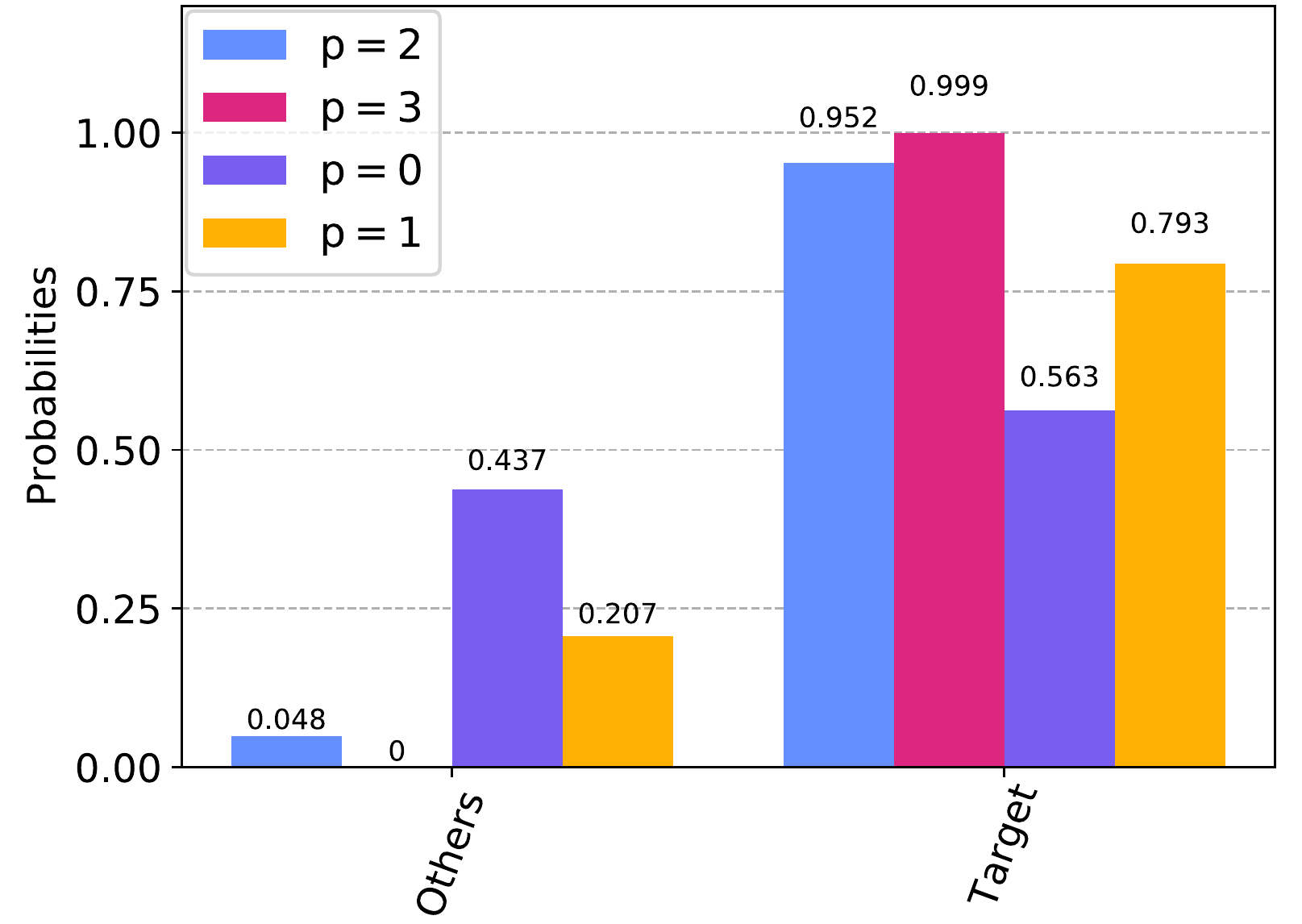}
    \caption{Result of Grover's search algorithm for GSA$(16,9)$ for different values of $p$. Here ``Target'' refers to the probabilities of all the target states combined, and ``Others'' refer to all the non-target states (the probabilities of all the non-target states combined).}
    \label{fig:outcome_9_targets}
\end{figure}

The output of the circuits are shown in figure \ref{fig:outcome_9_targets}. The success probabilities for $p=2,3$ are above the thresholds in table \ref{tab:optimal_k_9}. Moreover, it is evident that when $p=0$, the success probability is just $56.3\%$ which makes detection of targets almost as likely as that of non-targets. This example highlights the fact that choosing $p=0$ is not always good strategy, as discussed in \S\ref{sec:optimalk}.

\subsubsection{4-qubits 8 targets}
\begin{figure}
    \includegraphics[width=1\hsize]{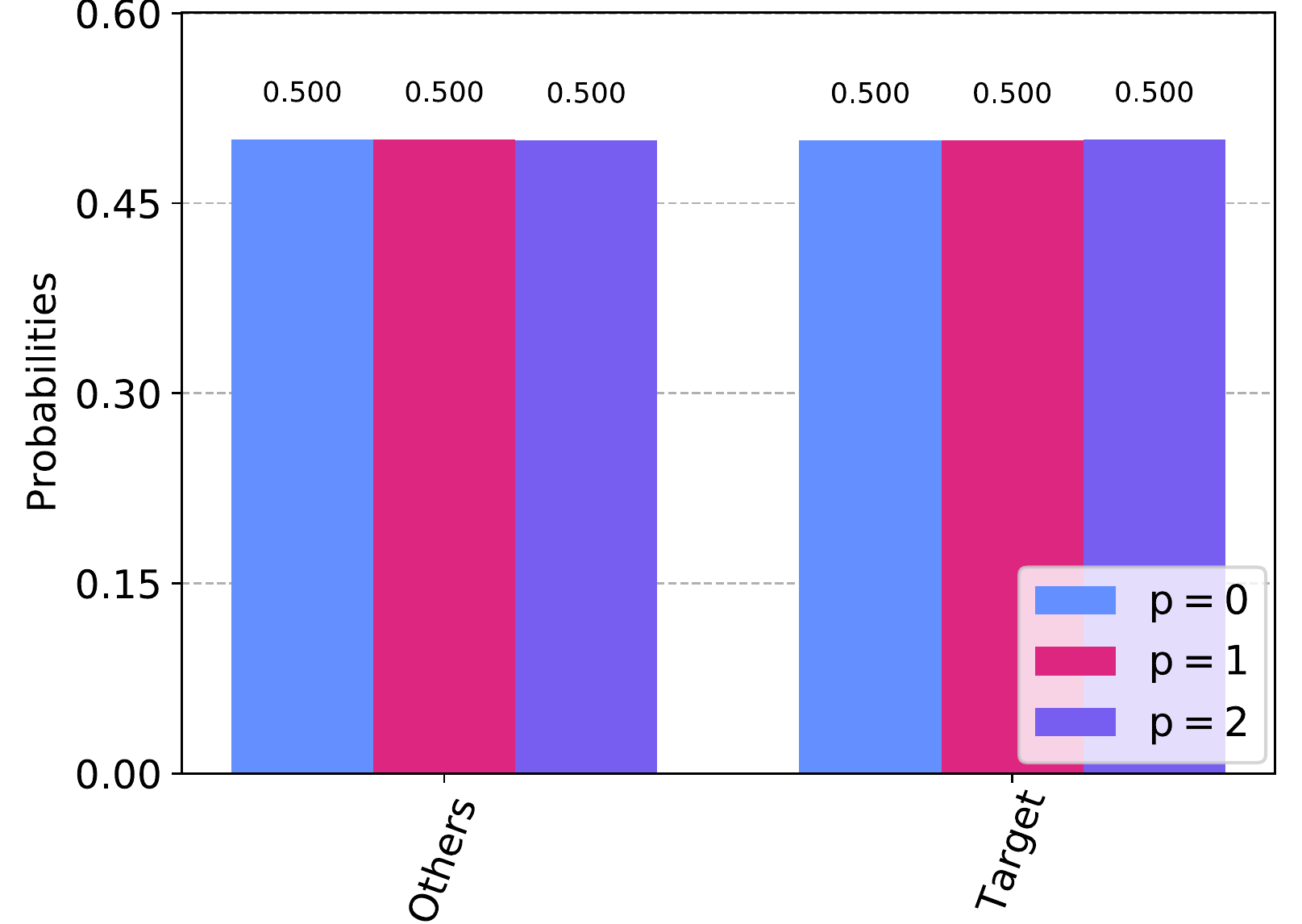}
    \caption{Result of Grover's search algorithm for GSA$(16,8)$ for different values of $p$. As the number of target states is half the size of the database, the search algorithm fails to amplify the target states.}
    \label{fig:outcome_4_8_targets}
\end{figure}

This example highlights the case of $M=N/2$ (as discussed in \S\ref{sec:optimalk}) for which, Grover's algorithm does not amplify the target states. The algorithm is executed for $p=0,1,2$ and figure \ref{fig:outcome_4_8_targets} shows that the success probability is 50\% for all three values of $p$. Therefore, increasing the number of iterations does not help.

\subsubsection{7-qubits 60 targets}
\begin{figure}
    \includegraphics[width=1\hsize]{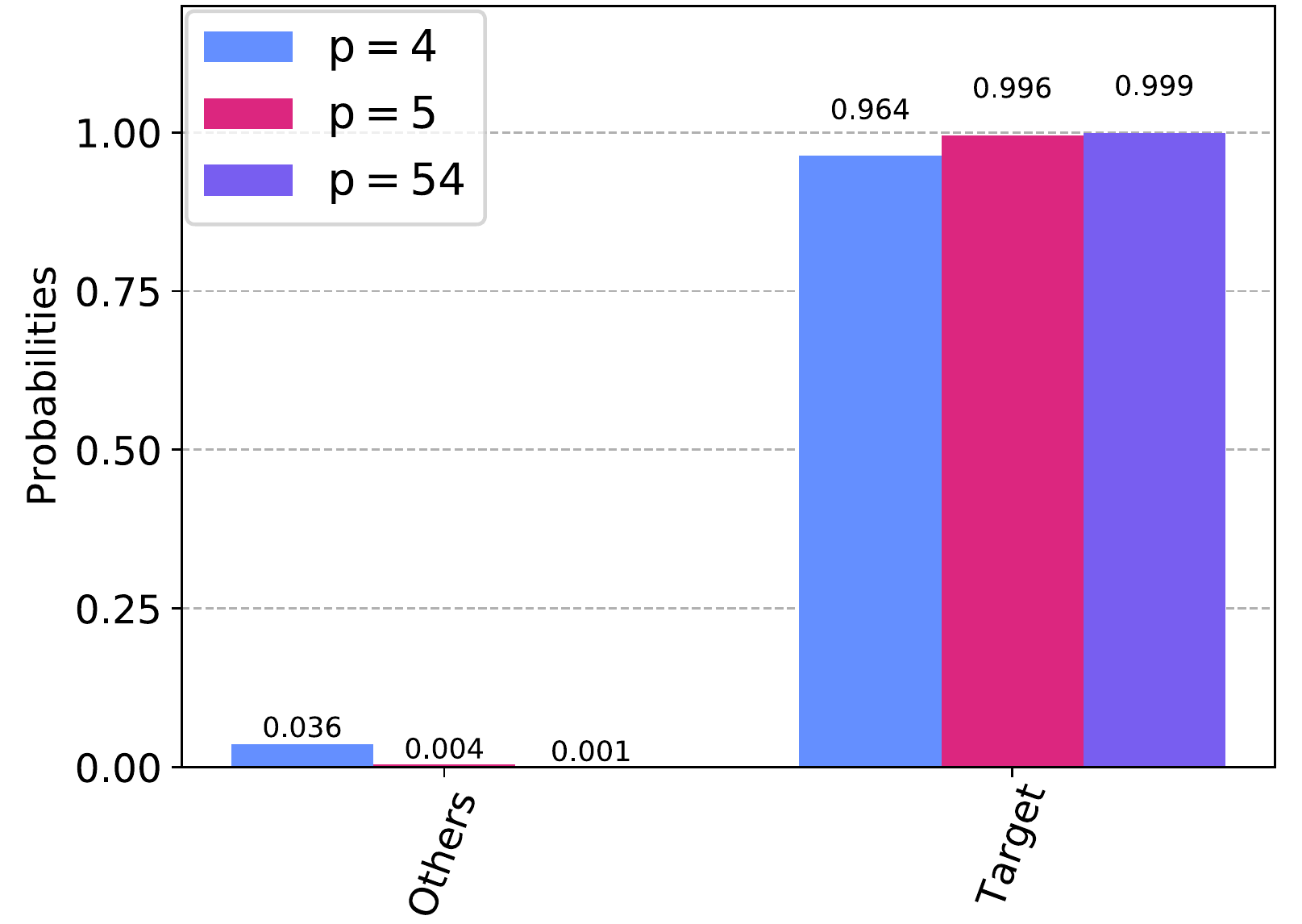}
    \caption{Result of Grover's search algorithm for GSA$(128,60)$ for different values of $p$. Here ``Target'' refers to the probabilities of all the target states combined, and ``Others'' refer to all the non-target states (the probabilities of all the non-target states combined).}
    \label{fig:outcome_60_targets}
\end{figure}

To further demonstrate higher dimensional Grover's search algorithm, this example runs the circuit for GSA$(128,60)$, with
\begin{align}
    \theta = \sin^{-1}\sqrt{\frac{60}{128}}.
\end{align}
{\renewcommand{\arraystretch}{1.5}
\begin{table}[H]
    \centering
    \begin{tabular}{>{\centering\arraybackslash}m{0.2\hsize} >{\centering\arraybackslash}m{0.2\hsize} >{\centering\arraybackslash}m{0.2\hsize}}
         $\delta$ & $p$ & $k_p$ \\
         \hline
         0.95 & 4 & 9 \\
         0.99 & 5 & 11 \\
         0.999 & 54 & 113
    \end{tabular}
    \caption{The value of $p$ and $k_p$ for a given success threshold $\delta$ for GSA$(128,60)$.}
    \label{tab:optimal_k_7_60}
\end{table}}
Table \ref{tab:optimal_k_7_60} shows the success probability thresholds and the corresponding values of $p$ and $k_p$, and figure \ref{fig:outcome_60_targets} shows the outputs. Note that, for $\delta = 0.999$, $p=54$ ($113$ iterations), which is a significant increase from $p=5$ ($11$ iterations) for $\delta=0.95$. Therefore, for many practical purposes, choosing $p=5$ is beneficial as on a physical quantum computer, the longer circuit suffers more decoherence, which may have adverse effects on the success probability.

\section{Conclusion and discussion}
A quantum circuit for Grover's search algorithm can be implemented for any number of qubits and any number of targets (irrespective of the choice of targets) by using the scheme presented in \S\ref{sec:circuit_construction}. However, choosing the optimal number of iterations of the combined operation $D U_\omega$ is critical to getting the desired success probability of search. Although the required number of iterations scale as $\mathcal{O}(\sqrt{N})$ for large $N$, for finite size of database, the optimal number of iterations may not follow that trend, especially when there are multiple targets to be searched for. The scheme in \S\ref{sec:optimalk}, solves the problem of determining the optimal number of iterations according to the threshold set on the success probability of the search algorithm. This scheme is crucial even for large $N$ if the number of search targets, i.e., $M$ is such that $\sqrt{M/N}$ is not small enough. In general, there is an upper-bound on the success probability that can be achieved in Grover's search, depending on the values of $N$ and $M$. In particular, for $M=N/2$, Grover's search algorithm does not amplify the target states at all. 

An extension/modification of Grover's algorithm can be explored. In a typical instance of Grover's algorithm with multiple targets, the amplified detection probabilities of all the target states, at the output, are equal. A notable work on the application of Grover's algorithm with non-uniform input distribution is \cite{biron1998generalized}. It is interesting to see if the algorithm can be modified such that the target states are amplified but the output has a given non-uniform probability distribution.

\section{Additional material}\label{sec:additional}
The examples presented above, are generated using the QASM simulator which is a part of IBM Qiskit. The python program used to assemble the circuits and execute them on the simulator, is available at \url{https://github.com/simanraj1123/n-qubit-Grover-s-search-on-IBM-Qiskit}
\nocite{*}
\bibliography{references}

\end{document}